\documentclass{article}

\usepackage{amsmath}
\usepackage{amssymb}
\usepackage{bm}
\DeclareMathSymbol{\hbar}{\mathord}{AMSb}{"7E}
\usepackage{mathptmx}
\usepackage{braket}
\usepackage{mathrsfs}
\usepackage{cite} 

\usepackage{graphicx}

\title{Polarization-controlled effective Rabi dynamics in driven Graphene: A Floquet–Magnus approach
}

\author{V. G. Ibarra-Sierra, J. L. Cardoso, C. Flores-Valente,\\
 A. Kunold and J. C. Sandoval-Santana$^{*}$
\\
\small
\'Area de F\'isica Te\'orica y Materia Condensada,\\
\small Universidad Aut\'onoma Metropolitana,\\
\small  Azcapotzalco, Av. San Pablo Xalpa 180, \\
 \small Ciudad de M\'exico, 02200, Ciudad de M\'exico,M\'exico\\
\small *jcss@azc.uam.mx\\
\small keywords: Floquet-Magnus expansion,\\
\small  Resonant driving, Dirac materials
}

\begin{document}


\maketitle

\begin{abstract}
Polarization ellipticity $\beta$ and the relative angle $\Delta$ between
electron momentum and driving field act as independent control parameters
for coherent dynamics in periodically driven Dirac systems. In this work,
we analyze the dynamics of resonantly driven Dirac electrons in graphene
under elliptically polarized electromagnetic radiation using the
Floquet–Magnus expansion.
Working in the interaction picture and applying a rotating-wave-type
transformation, we derive an effective two-level Hamiltonian that governs
the macromotion at resonance ($\omega = \Omega/2$). The resulting
quasienergy splitting depends nontrivially on $\beta$ and $\Delta$ through
interference between the Bessel harmonics $J_0(\zeta)$ and $J_2(\zeta)$.
Circular polarization ($\beta = \pm 1$) restores rotational symmetry and
yields a $\Delta$-independent effective Rabi frequency, whereas elliptical
and linear polarizations produce anisotropic responses with a
$\pi$-periodic angular modulation.
Beyond spectral properties, we identify a polarization-induced phase that
acts as an effective initial Floquet kick, shifting the effective initial
conditions and producing measurable shifts in the timing of occupation
oscillations, whose sign depends on both helicity and relative orientation.
Through an explicit Fourier decomposition of the time-evolution operator,
we separate macromotion from micromotion contributions and validate the
zeroth-order Magnus approximation via numerical simulations, achieving
root-mean-square errors of $\sim 1\%$ over 100 driving periods in the
weak-field regime.
These results establish polarization ellipticity and relative orientation
as tunable and experimentally accessible knobs for quantum control in
two-dimensional Dirac materials, with direct implications for
time-resolved spectroscopy.
\end{abstract}

\section{Introduction}

The ability to engineer the properties of quantum materials through periodic driving has emerged
as one of the most promising directions in condensed-matter physics. The Floquet formalism
provides a rigorous framework to describe systems whose Hamiltonians are periodic in time,
mapping the full time-dependent problem onto an effective static description in terms of
quasienergies and Floquet states~\cite{Goldman2014, Eckardt2015, Bukov2015,sandoval2019method}. The central promise
of Floquet engineering is that the properties of the effective Hamiltonian governing the
long-time dynamics can be systematically designed by tuning the parameters of the driving field,
enabling the coherent control of electronic phases that have no equilibrium
counterpart~\cite{Oka2019}. In addition to  the effective Hamiltonian, the time-evolution operator itself can acquire additional phase factors  known as Floquet kicks  that reshape the initial conditions of the driven system and leave measurable imprints on the transient dynamics \cite{Goldman2014, Eckardt2015, Bukov2015}.

Two-dimensional (2D) materials combine exceptional physical and chemical properties with a rich
landscape of novel quantum phases~\cite{novoselov2012roadmap, peng2016electronic,
wehling2014dirac, ferrari2015science, Mehboudi5888Strain, Villanova2016Spin}, making them
ideal platforms for ultrafast materials science, where optical and mechanical control of solids
demands theoretical tools beyond conventional approaches~\cite{naumis2017electronic}. Under
electromagnetic radiation, electrons become dressed and band structures are modified~\cite{
lopez2010graphene, kibis2010metal, Foa2011Tuning, kristinsson2016control,Kunold2020,Mojarro2020Kek, Sandoval2020}, giving rise to gaps, nonlinear effects, and enhanced photocurrents~\cite{sun2012ultrafast,
Chan2017Photocurrents}. The Floquet
framework~\cite{Oka2019Floquet} provides the natural language for these analyses, systematically exploiting
time-periodicity to uncover complex topological phase diagrams~\cite{Roman2017TopologicalEdge} through advanced mathematical
techniques~\cite{Goldman2014,Eckardt2015,sandoval2019method}.

Graphene, with its linear Dirac dispersion and massless charge carriers, occupies a privileged
position in this program. Graphene has served as the primary testing ground for light-induced
topological phenomena: photoinduced band-gap opening \cite{Oka2019Floquet,Oka2009}, anomalous Hall responses \cite{Wang2013}, and the
generation of Floquet sidebands have all been predicted and, more recently, directly observed
through time-resolved photoemission spectroscopy~\cite{McIver2020, Merboldt2025}. These
experimental advances confirm that Floquet engineering in graphene is realizable even in the
presence of fast decoherence, opening the door to a systematic exploration of the resonant
regime that had remained theoretically underexplored.

Beyond the off-resonant dressing-field limit, where high-frequency expansions are well controlled and the effective Hamiltonian can be derived perturbatively in $1/\Omega$, the resonant regime
presents qualitatively distinct physics. When the driving frequency matches an interband transition
energy, the Floquet harmonics cannot be treated as small corrections and a different theoretical
strategy is required. In this context, the combination of the interaction picture with a
rotating-wave-type transformation provides a natural starting point, reducing the full periodically
driven problem to a generalized Rabi model in which the effective coupling between Floquet-dressed
bands is explicit and directly controllable.

A feature common to all these prior analyses is either the restriction to a specific polarization
state, typically circular or linear, or the treatment of polarization effects as secondary. Yet
polarization is among the most experimentally accessible knobs: the ellipticity of a radiation
field can be continuously tuned from circular ($\beta = \pm 1$) to linear ($\beta = 0$), and this
variation profoundly alters the symmetry of the light-matter interaction. The role of the electromagnetic polarization state in shaping the Floquet spectrum has been
investigated in off-resonant contexts for tilted Dirac semimetals~\cite{Ibarra2019, Kunold2020, Sandoval2020},
where elliptically polarized light was shown to dynamically tune band-gap openings and topological
transitions. In the resonant regime, however, the interplay between polarization, resonance
condition, and quasienergy splitting has not been analyzed within a unified framework that is
simultaneously exact, analytically tractable, and extendable to general ellipticity. The
Floquet--Magnus expansion offers precisely such a framework: by expressing the evolution operator
as the exponential of a Magnus operator and truncating at finite order, one obtains a stroboscopic
effective Hamiltonian that captures the macromotion while the micromotion operator accounts for the fast intra-period
oscillations~\cite{Goldman2014, Eckardt2015}. One of the key advantages of the Magnus expansion is that, at any order of truncation,  the time-evolution operator strictly preserves unitarity. Moreover, it has been shown to  provide highly accurate results in the weak-field regime, where the driving perturbation 
is small \cite{Magnus1954,Blanes2010,Fernandez2005}. 

In this work, we present a complete Floquet-Magnus analysis of resonantly driven Dirac electrons
in graphene subject to elliptically polarized electromagnetic radiation of arbitrary ellipticity
$\beta$. Working in the interaction picture, we apply a unitary transformation analogous to a
rotating-wave approximation but exact in its handling of the diagonal modulation to isolate the
effective coupling responsible for interband transitions. Applying the Jacobi-Anger expansion to
the resulting time-dependent matrix elements, we derive the full Fourier decomposition of the
interaction Hamiltonian, from which the resonance condition $\omega = \Omega/2$ emerges naturally
as the condition for a single dominant Floquet harmonic to survive the stroboscopic time-average.
The resulting effective two-level Hamiltonian depends nontrivially on both $\beta$ and the
relative angle $\Delta$ between the electron momentum and the polarization direction, through a
pair of Bessel functions whose interference governs the angular structure of the quasienergy
splitting. Beyond the quasienergy spectrum, we show that the polarization state induces a
time-independent phase factor in the evolution operator that acts as an initial Floquet kick,
effectively redefining the initial conditions of the driven system and producing a measurable time
shift in the occupation probability oscillations. To characterize the full dynamics, we develop a
Fourier decomposition of the evolution operator that explicitly separates macromotion from
micromotion contributions. Numerical simulations validate the zeroth-order Floquet--Magnus
approximation across a wide parameter range in the weak-driving limit, while
deviations at longer times are shown to be consistent with secular accumulation of higher-order
Magnus corrections.

The paper is organized as follows. Section \ref{sec:model} introduces the
graphene Hamiltonian under polarized electromagnetic radiation and defines the vector potential
parametrization. Section \ref{sec:interaction} formulates the interaction picture and applies the
unitary transformation that generates the generalized Rabi model. Section \ref{sec:floquet}
derives the stroboscopic effective Hamiltonian via the Floquet--Magnus expansion and establishes
the resonance condition. Section \ref{sec:micromotion} develops the macro/micromotion
decomposition using a Fourier expansion of the kick operator. Section \ref{sec:results} presents and discusses
the numerical and analytical results for the quasienergy splitting, the polarization-induced phase,
and the occupation probability dynamics. Section \ref{sec:conclusions} collects our conclusions
and outlines extensions of the methodology to other Dirac materials and driving regimes.

\begin{figure}[b]
\begin{center}
\includegraphics[scale=0.49]{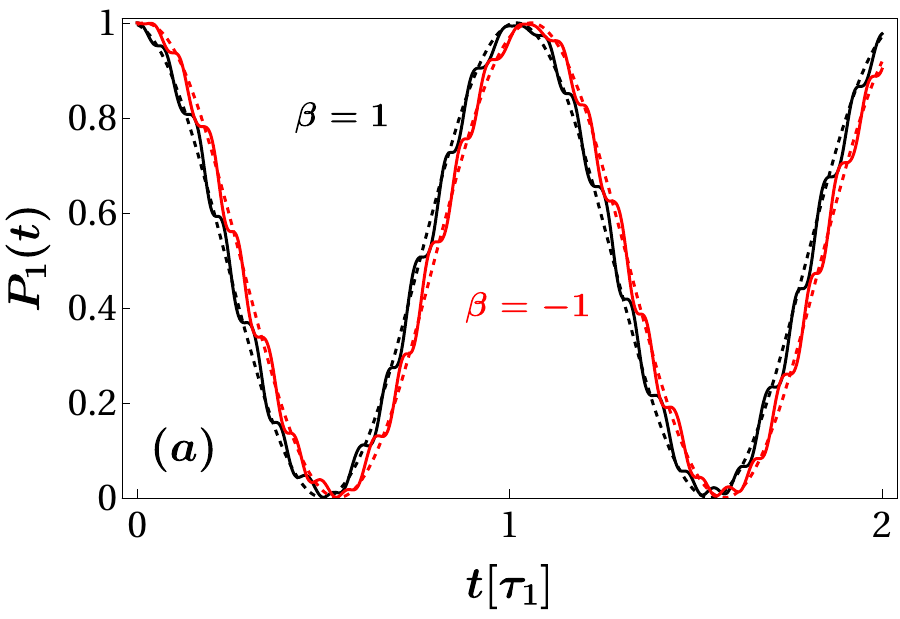}
\includegraphics[scale=0.49]{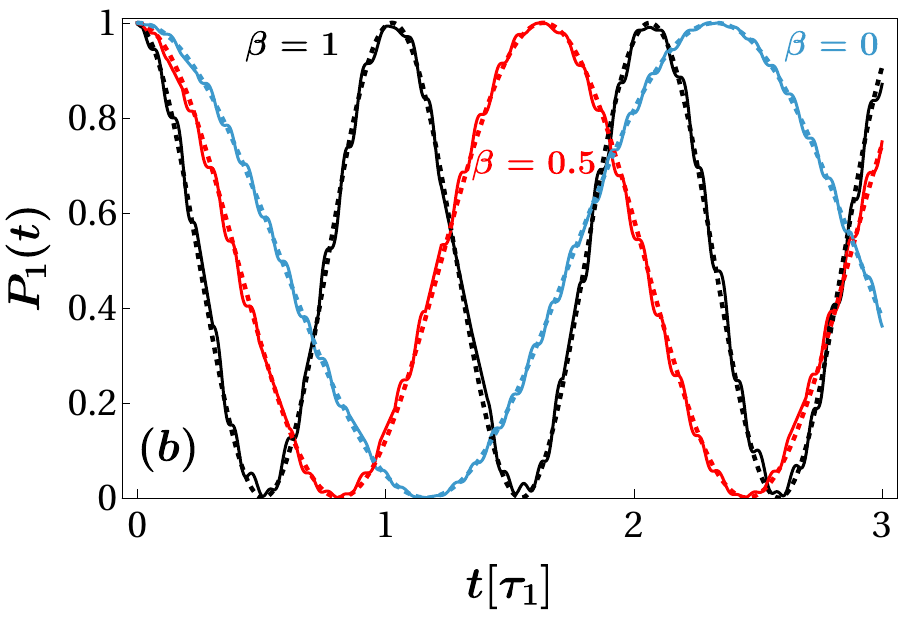}
\end{center}
\caption{\label{Fig1} Time evolution of the population of the
valence band $P_1(t)$ in the resonant regime $\omega = \Omega/2$.
(a) Comparison between the full numerical solution (solid lines)
and the analytical solution describing the macromotion (dashed lines)
for opposite polarization chiralities, $\beta = \pm 1$, with $\Delta=\pi/4$.
The slight horizontal displacement reflects the phase-induced initial Floquet kick.
(b) Numerical and analytical evolution of $P_1(t)$
for $\beta = 1,\,0.5,\,0$ with $\Delta=\pi/7$,
demonstrating the progressive modification of the effective Rabi
frequency as $\beta$ is reduced.
The envelope dynamics is governed by the effective Hamiltonian
and the initial Floquet kick, while small deviations originate from micromotion.
}
\end{figure}

\begin{figure}[b]
\begin{center}
\includegraphics[scale=0.6]{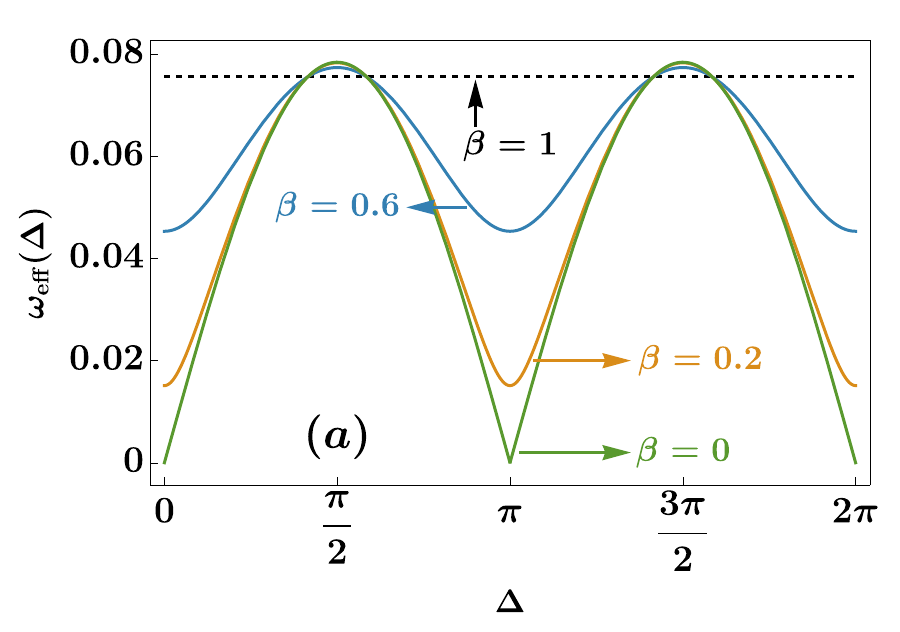}
\includegraphics[scale=0.5]{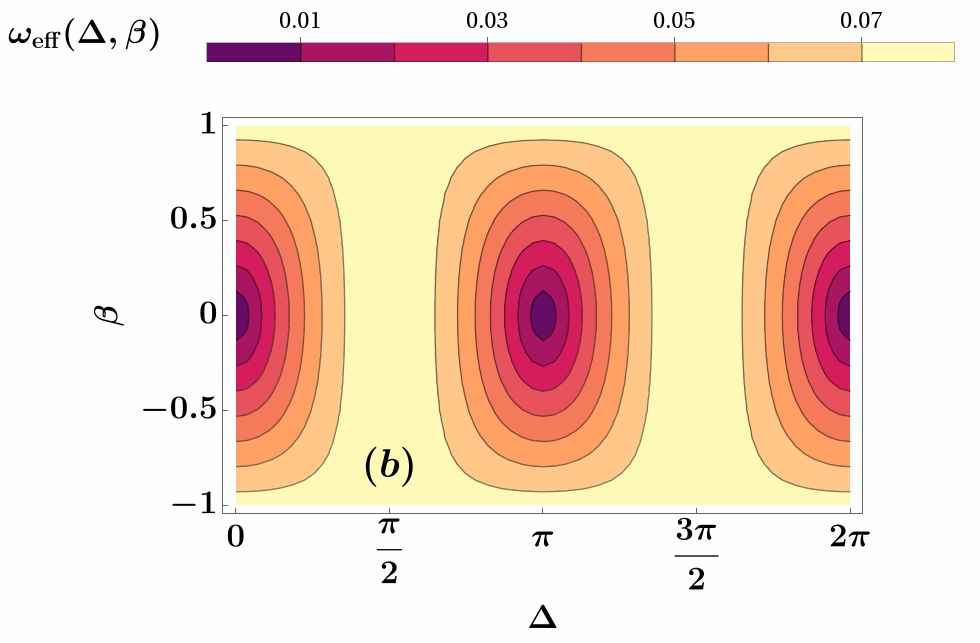}
\end{center}
\caption{\label{Fig2} Effective Rabi frequency $\omega_\mathrm{eff}$ as a
function of the polarization parameter $\beta$ and the relative angle
$\Delta$ between the polarization direction and the electron momentum.
(a) Dependence of $\omega_\mathrm{eff}(\Delta)$ for $\beta=1,\,0.6,\,0.2,\,0$.
Note that circular polarization ($\beta=1$) restores rotational symmetry,
yielding an angle-independent effective Rabi frequency,
whereas decreasing $\beta$ enhances the anisotropic modulation.
For $\Delta = n\pi$, where $n$ is an integer, and $\beta=0$, 
one has $\omega_{\mathrm{eff}}=0$, and the dynamics is therefore
completely governed by micromotion.
(b) Contour plot of $\omega_{\mathrm{eff}}(\Delta,\beta)$,
exhibiting a $\pi$-periodic angular structure and a progressive increase
in anisotropy as the polarization approaches the linear limit ($\beta=0$).
}
\end{figure}

\begin{figure}[b]
\begin{center}
\includegraphics[scale=0.6]{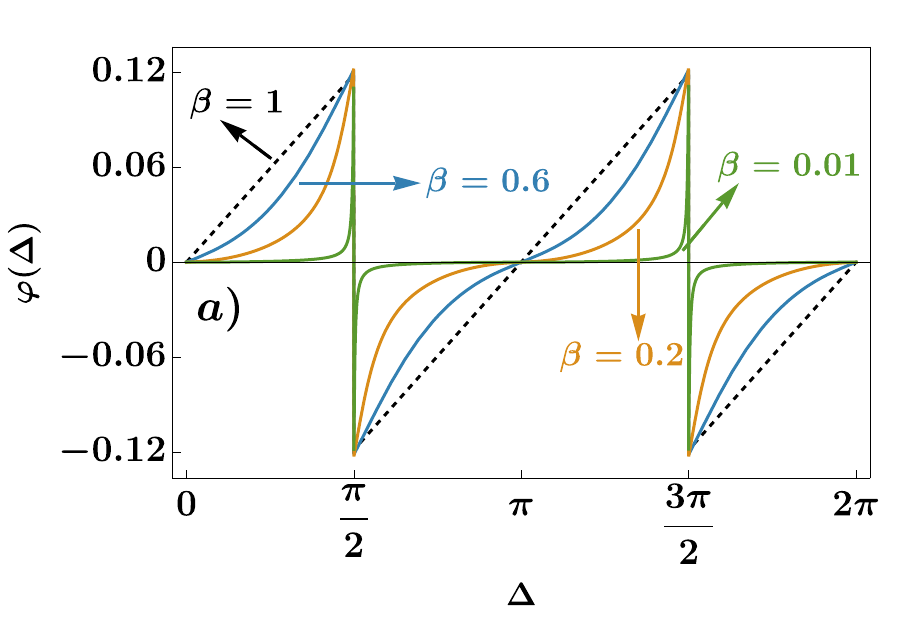}
\hfill
\includegraphics[scale=0.5]{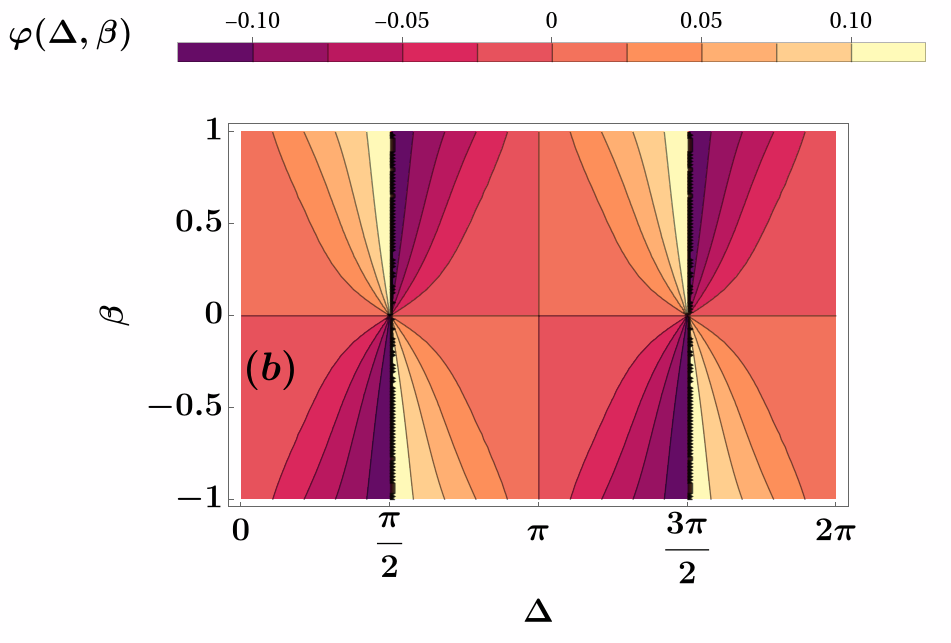}
\end{center}
\caption{\label{Fig3}  Phase induced by the initial Floquet kick $\varphi$ as a function of 
$\beta$ and $\Delta$. (a) Angular dependence of $\varphi(\Delta)$ for 
$\beta = 1,\, 0.6,\, 0.2,\, 0.01$. For all values of $\beta$, the 
phase undergoes an abrupt sign reversal at $\Delta = \pi/2$ and 
$3\pi/2$. In the circular limit ($\beta = 1$, dashed line), $\varphi$ 
varies linearly with $\Delta$ within each half-period. As $\beta$ 
decreases, the linear behavior is lost and the phase develops an 
increasingly asymmetric profile, reflecting the enhanced angular 
sensitivity of the effective coupling. The anisotropy is most pronounced for 
small $\beta$ and gradually vanishes as the polarization approaches the circular limit ($\beta \rightarrow 1$). (b) Contour plot of $\varphi(\Delta,\beta)$ 
showing a $\pi$-periodic angular structure with alternating positive 
and negative regions separated by nodal lines at 
$\Delta = \pi/2,\, 3\pi/2$.
}
\end{figure}

\section{Graphene under polarized electromagnetic radiation}\label{sec:model}

To analyze the driven dynamics,
we introduce the theoretical framework used to describe Dirac electrons in graphene subject to polarized electromagnetic radiation.
The low-energy Hamiltonian of graphene \cite{torres2014introduction,katsnelson2012graphene}
is given by
\begin{equation}\label{ec:LowEnergy_H}
\hat{H}^{0}_{\xi}=\hbar v_F \bm{k} \cdot \bm{\hat \sigma}, 
\end{equation}
where  $v_F\approx 10^{6}$ m/s is the Fermi velocity, $\bm{k}=(k_x,k_y)$ the momentum vector  and $\bm{\hat \sigma}=(\xi \hat{\sigma}_x,\hat{\sigma}_y)$ with $\xi =\pm 1$ refers to the \textbf{K} and  \textbf{K}' valleys. The pseudo-spin operators $\hat{\sigma}_x$ and $\hat{\sigma}_y$ are the Pauli matrices. The electronic band structure is characterized by two Dirac cones, where the conduction band (CB) is described by $E_{1}=v_F\hbar|\boldsymbol{k}|$ and $E_{2}=-v_F\hbar|\boldsymbol{k}|$ corresponds to the valence band (VB).  The eigenfunctions for each band are given by 
\begin{equation}
\ket{\psi_{\boldsymbol{k},\xi,1}}=\frac{1}{\sqrt{2}}\Big[\xi \mathrm{e}^{-i\xi\theta_{\boldsymbol{k}}}\ket{A} + \ket{B}\Big]    \qquad   \ket{\psi_{\boldsymbol{k},\xi,2}}=\frac{1}{\sqrt{2}}\Big[\xi \mathrm{e}^{-i\xi\theta_{\boldsymbol{k}}}\ket{A} - \ket{B}\Big]
\end{equation}
where $\theta_{\boldsymbol{k}}=tan^{-1}(k_y/k_x)$,
$\left| A \right\rangle = \left(1,0\right)^{T}$ and $\left| B \right\rangle = \left(0,1\right)^{T}$ are the spinors that describe the sublattice degree of freedom ($\mathrm{A}$ and $\mathrm{B}$).

To analyze electron dynamics in graphene under electromagnetic radiation,
we apply the Peierls substitution
$\boldsymbol{k} \to \boldsymbol{k} - e\boldsymbol{A}/\hbar$ \cite{dey2018photoinduced} 
to the low-energy Hamiltonian in Eq.~(\ref{ec:LowEnergy_H}),
where $\boldsymbol{A} = (A_x(t), A_y(t))$ is the vector potential.
For this problem, we assume normal incidence and choose a gauge
in which $A_x(t)$ and $A_y(t)$ depend only on time. We then obtain
\begin{equation}\label{eq:Ht}
\hat{H}_{\xi}(t)= \hbar v_F\left(\boldsymbol{k} - \frac{e}{\hbar} \boldsymbol{A}\right)\cdot\boldsymbol{\hat \sigma}.
\end{equation}
The vector potential is thus given by
\begin{equation}
\boldsymbol{A}=\frac{E_0}{\Omega} \left( \cos \theta_p \cos \Omega t - \beta \sin \theta_p \sin \Omega t ,  \sin \theta_p \cos \Omega t + \beta \cos \theta_p \sin \Omega t \right),
\end{equation}
where $E_{0}$ is the electric field amplitude, $\Omega$ is the angular frequency,
and $\theta_{p}$ is the initial polarization angle measured from 
the $A_{x}^{+}$ axis.
The parameter $\beta$ characterizes the polarization and helicity
of the wave: $-1 \leq \beta < 0$ corresponds to negative (clockwise) helicity,
$0 < \beta \leq 1$ to positive (counterclockwise) helicity;
$\beta = 0$ yields linear polarization, $|\beta|=1$ circular polarization,
and $0<|\beta|<1$ elliptical polarization.
The eccentricity of the polarization ellipse as a function of $\beta$
is given by $\varepsilon = (1 - \beta^2)^{1/2}$.

\section{Hamiltonian in the interaction picture}\label{sec:interaction}
To facilitate the treatment of the time-dependent dynamics, we rewrite the time-dependent Hamiltonian in the interaction picture.
We take the time-dependent Hamiltonian given in Eq. (\ref{eq:Ht})
and formulate the corresponding Schrödinger equation in the
interaction picture\cite{sakurai1995modern} given by
\begin{equation}\label{eq:DiracEquationInteraction}
i\frac{d}{dt}\ket{\boldsymbol{\chi}(t)}=\hat{H}_{I,\xi}(t)\ket{\boldsymbol{\chi}(t)},
\end{equation}
where $\boldsymbol{\chi}(t)= \sum_{\mu} \chi_{\mu}(t)\ket{\mu}$ collects the wavefunction components
in the
conduction band (CB) and valence band (VB) within the interaction picture.
The coefficients are given by
$\chi_{\mu}(t)=\exp\left(i E_\mu t/ \hbar\right)\braket{\psi_{\boldsymbol{k},\mu}|\boldsymbol{\Psi}(t)}$
and $\ket{\boldsymbol{\Psi}(t)}$ denotes the time-dependent two-component spinor in the Schrödinger picture.
The subscript $\mu = 1, 2$ labels the CB and VB, respectively.
The Hamiltonian in the interaction picture is given by $\hat{H}_{I,\xi}(t)$.
When expressed in the CB–VB basis, its matrix elements are
$[\hat{H}_{I,\xi}(t)]_{\mu,\nu}=\exp\left[i (E_\mu-E_\nu)t/\hbar\right]\bra{\psi_{\boldsymbol{k},\mu}}\hat{V}(t)\ket{\psi_{\boldsymbol{k},\nu}}$. Expressed in the basis of the operators $\hat{\sigma}_+$, $\hat{\sigma}_-$, and $\hat{\sigma}_z$, the Hamiltonian takes the form
\begin{equation}\label{eq:Hinter}
\hat{H}_{I,\xi}(t)=
 R(t)\hat{\sigma}_z- i \xi Q(t) e^{2i \xi \omega t} \hat{\sigma}_{+}+i \xi Q(t) e^{-2i \xi \omega t} \hat{\sigma}_{-}
\end{equation}
where
\begin{eqnarray}
    R(t) &=&\lambda(\beta 
    \sin{(\Omega t)}\sin{(\Delta)}- \cos{(\Omega t)}\cos{(\Delta)}),\\
    Q(t) &=& \lambda(\beta 
    \sin{( \Omega t)}\cos{(\Delta)}+ \cos{( \Omega t)}\sin{(\Delta)}),
\end{eqnarray}
$\lambda=e v_{F} E_0/\hbar\Omega$ and $\Delta=\theta_p - \theta_k$ denotes the relative angle between the polarization direction and the electron momentum. The operators $ \hat \sigma_+=(\hat{\sigma}_x+i\sigma_y)/2$ and $\hat \sigma_-=(\hat{\sigma}_x-i\sigma_y)/2$ act as raising and lowering operators in the pseudospin basis.
In this definition, we have introduced a set of renormalized momenta,
$\kappa_x = (v_F/\Omega)k_x$ and $\kappa_y = (v_F/\Omega)k_y$, which are related to the frequency through
$\omega = \Omega (\kappa_x^2 + \kappa_y^2)^{1/2}$.

To analyze this system in resonance regime, we introduce the unitary transformation
\begin{equation}\label{eq:Uz}
\hat U_z(t)=\exp\!\left[-if(t)\hat\sigma_z\right]
\end{equation}
to eliminate the time-dependent diagonal component  $R(t)\hat\sigma_z$ of the interaction Hamiltonian Eq.(\ref{eq:Hinter}). 
Since the operators satisfy the commutation relations $
[\hat\sigma_z,\hat\sigma_\pm]=\pm 2\hat\sigma_\pm $
the transformation acts nontrivially only on the off-diagonal operators, generating phase factors according to
\begin{equation}
\hat U_z^\dagger(t) \hat\sigma_\pm \hat U_z(t)
= e^{\pm 2 i  \int^t f(t_1) dt_1} \hat\sigma_\pm \, ,
\end{equation}
which were obtained using the Lie algebra method \cite{sandoval2019method}. Applying the unitary transformation $\hat{U}^\dagger_z(t)$ from the left to Eq.(\ref{eq:DiracEquationInteraction}),  considering Eq.(\ref{eq:Hinter}) and introducing the definitions
$\ket{\boldsymbol{\chi}(t)}=\hat{U}_z(t)\ket{\boldsymbol{\phi}(t)}$ and $f(t)=R(t)$, we obtain
\begin{multline}\label{eq:phiInt}
i\frac{d}{dt}\ket{\boldsymbol{\phi}(t)}=\bigg(-i \xi Q(t) e^{2i \xi \omega t}  e^{i2 \int R(t_1) dt_1}\hat{\sigma}_{+}\\
+i \xi \ Q(t) e^{-2i  \xi \omega t} e^{-i2 \int R(t_1) dt_1} \hat{\sigma}_{-}  \bigg)\ket{\boldsymbol{\phi}(t)}\,.
\end{multline}
The resulting transformed Hamiltonian contains only the operators 
$\hat{\sigma}_\pm$ dressed by time-dependent functions. This procedure 
can be interpreted as moving to a time-dependent rotating frame generated 
by $\hat{\sigma}_z$, in which the longitudinal modulation associated with 
$R(t)$ is absorbed into the phase of the transverse couplings, isolating 
the dynamical processes responsible for interband transitions. The 
resulting Hamiltonian takes the form of a generalized Rabi-type model 
whose matrix elements are complex and time-dependent.

Using trigonometric identities, $Q(t)$ and $\int R(t) dt$ can be written as
\begin{eqnarray}
Q(t) &=& \lambda \gamma \sin(\Omega t + \eta),\\
\int_0^t R(t_1) dt_1 &=& -\frac{\lambda \delta}{\Omega} \sin(\Omega t + \rho)+cte,
\end{eqnarray}
where
\begin{align}
& \gamma = \sqrt{\sin^2\Delta + \beta^2 \cos^2\Delta} 
\,\,\,\,\,\, ,&
\tan \eta = \frac{\tan\Delta}{\beta} \,\,\,\,\,\, ,
\label{eq:functions}\\
& \delta = \sqrt{\beta^2 \sin^2\Delta + \cos^2\Delta}\,\,\,\,\,\, ,
& \tan\rho = \beta \tan\Delta \,\,\,\,\,\, .
\end{align}
Using these definitions, Eq.~(\ref{eq:phiInt}) takes the form
\begin{equation}
i\frac{d}{dt}\ket{\boldsymbol{\phi}(t)}=\big[-i \xi \lambda \gamma \sin \left( \Omega t + \eta \right) e^{2i \xi \omega t}  e^{-i \frac{2 \lambda \delta}{\Omega}\sin \left( \Omega t + \rho \right) }\hat{\sigma}_{+}+ H.c. \big]\ket{\boldsymbol{\phi}(t)}\, .
\end{equation}
Applying the Jacobi-Anger expansion $e^{i z \sin \theta}
= \sum_{n=-\infty}^{\infty} J_n(z)\, e^{i n \theta}$  and defining $t' = \Omega t + \rho$, we obtain
\begin{equation}\label{eq:EcuShroInter}
i\frac{d}{dt^{\prime}}\ket{\boldsymbol{\phi}(t^\prime)}=\hat H_{I,\xi}(t^\prime) \boldsymbol{\phi}(t^\prime)\, ,
\end{equation}
where
\begin{equation}\label{eq:HinFourier}
\hat H_{I,\xi}(t^\prime) =\sum_n \big[ C_n e^{-i t^{\prime} \left( n- \frac{2 \omega}{\Omega}  \right)} \hat \sigma_+ + H.c.  \big] \, .
\end{equation}
This Hamiltonian is periodic in $t$ with 
period $\tau=2\pi/\Omega$.
The corresponding Fourier coefficients are given by
\begin{equation}
C_n=-i \xi \frac{ \lambda \gamma}{2\Omega} e^{-i 2 \xi \frac{\omega}{\Omega}\rho} \big[ J_{n-1}(\zeta) e^{-i (\rho-\eta)}- J_{n+1}(\zeta) e^{i (\rho-\eta)}\big]\, .
\end{equation}
where $\zeta=2\lambda \delta/\Omega$.
The structure of the interaction Hamiltonian  $\hat H_{I,\xi}(t^\prime)$
is particularly revealing because it highlights several key physical features.
First, the relative weight of the multiphoton channels is controlled by
the Bessel functions $J_{n\pm1}(\zeta)$, which determine the intensity of
each Floquet harmonic.
In the weak-field limit, $\zeta \ll 1$, the dominant contribution
comes from the zeroth-order term,
since $J_{0}(\zeta)\sim 1$ near $\zeta=0$,
whereas $J_{n\ne 0}(\zeta)\rightarrow 0$;
this is the perturbative regime, in which only a few harmonics contribute.
By contrast, when $\zeta \gtrsim 1$, several channels become important,
signaling a nonperturbative regime. Second, the angular and polarization
dependence is entirely encoded through the parameters
$\gamma$, $\rho$, $\delta$, and $\eta$, so that all geometric
information associated with the angle $\Delta$ and the
polarization parameter $\beta$ is captured by these quantities.

\section{Stroboscopic Effective Hamiltonian from Floquet–Magnus Theory}\label{sec:floquet}
To obtain an effective description of the driven dynamics,
we now apply Floquet–Magnus theory to the interaction-picture Schrödinger equation.
We consider the Schrödinger equation in the interaction picture in $\phi$ representation,
Eq.~(\ref{eq:EcuShroInter}), whose formal solution can be written as
\begin{equation}\label{eq:Uphii}
|\phi(t')\rangle = \hat U_\phi(t') |\phi_0\rangle,
\qquad
\hat U_\phi(t') = e^{\hat\Omega(t')},
\end{equation}
where $\hat\Omega(t')$ is the Magnus operator
\cite{Magnus1954,Blanes2010,Fernandez2005} expressed as an infinite
series of nested commutators,
\begin{equation}
\hat\Omega(t') = \sum_{n=1}^{\infty} \hat\Omega_n(t').
\end{equation}

The first  three terms in the Magnus expansion are
\begin{align}
\hat\Omega_1(t') &= -i \int_0^{t'} \hat H_{I,\xi}(t_1)\, dt_1,
   \label{eq:magnus01}\\
\hat\Omega_2(t') &= -\frac{1}{2} \int_0^{t'} dt_1
\int_0^{t_1} dt_2
\,[\hat H_{I,\xi}(t_1), \hat H_{I,\xi}(t_2)], \\
\hat\Omega_3(t') &= \frac{i}{6}
\int_0^{t'} dt_1
\int_0^{t_1} dt_2
\int_0^{t_2} dt_3
\Big(
[\hat H_{I,\xi}(t_1),[\hat H_{I,\xi}(t_2),\hat H_{I,\xi}(t_3)]]\nonumber\\
&\,\,\,\,\,\,\,\,\,\,\,\,\,\,\,\,\,\,+[\hat H_{I,\xi}(t_3),[\hat H_{I,\xi}(t_2),\hat H_{I,\xi}(t_1)]]
\Big).\label{eq:magnus03}
\end{align}
For a periodically driven system with period $\tau=2\pi/\Omega$,
the zeroth-order effective Hamiltonian of Eq. \eqref{eq:magnus01}
is defined as the time average
over one period,
\begin{equation}
\hat H_{\mathrm{eff}}^{(0)}
= \frac{1}{\tau}\hat\Omega_1(\tau)
= -\frac{i}{\tau}\int_0^{\tau}\hat H_I(t_1)\,dt_1.
\end{equation}
Substituting the Fourier decomposition of the interaction Hamiltonian,
Eq. (\ref{eq:HinFourier},) into the definition of $\hat H_{\mathrm{eff}}^{(0)}$
and evaluating the time integral term by term, we obtain
\begin{equation}
\frac{1}{\tau}
\int_0^\tau e^{i (n\Omega - 2\omega)t_1}\, dt_1
=
\frac{1}{\tau}
\frac{e^{i (n\Omega - 2\omega)\tau}-1}
{i (n\Omega - 2\omega)},
\end{equation}
since $\tau = 2\pi/\Omega$, the numerator vanishes unless
\begin{equation}
n\Omega - 2\omega = 0.
\end{equation}
Therefore, the time average over one period eliminates all rapidly oscillating contributions and retains only the stationary terms satisfying the resonance condition
\begin{equation}
n\Omega = 2\omega, 
\end{equation}
which consequently survive in the effective description and govern the long-time macromotion.
The lowest nonvanishing  resonance corresponds to $n=1$, which provides the dominant 
contribution in the weak-driving regime ($\zeta\ll 1$).
Higher-order resonances  ($n>1$) are progressively suppressed
due to the rapid decay of the corresponding Fourier coefficients,
leading to the resonance condition
\begin{equation}\label{eq:res}
\omega = \frac{\Omega}{2}.
\end{equation}
Under the resonance condition Eq. (\ref{eq:res}) and considering $n=1$, the $C_n$ Fourier coefficients take the form
\begin{equation}\label{eq:C1}
C_1 =
-i\xi \frac{\gamma \lambda}{2\Omega }
e^{ -i \xi \rho} \left[
J_{0}\!\left( \zeta \right)
e^{ -i (\rho-\eta) }
-
J_{2}\!\left( \zeta \right)
e^{ i (\rho - \eta) }
\right].
\end{equation}
The effective Hamiltonian in the interaction picture in the resonance regime is therefore given by
\begin{equation}\label{eq:energy}
\hat{H}^{(0)}_{\mathrm{eff}} =
C_1 \hat{\sigma}_{+}
+
C_1^{*} \hat{\sigma}_{-},
\end{equation}
which corresponds to an effective two-level system.
The quasienergy spectrum follows immediately as
\begin{equation}
\varepsilon_{eff,\pm}= \pm |C_1|= \pm 
\frac{\gamma \lambda}{2 \Omega}
\sqrt{
J_0^2\!\left( \zeta \right)
+
J_2^2\!\left( \zeta \right)
-
2 J_0\!\left( \zeta \right)
J_2\!\left( \zeta \right)
\cos(2(\rho - \eta))
},
\end{equation}
and the corresponding quasi-eigenstates can be written as
\begin{equation}
\ket{u_{\pm}} =
\frac{1}{\sqrt{2}}
\begin{pmatrix}
e^{ i \theta } \\
\pm e^{- i \theta }
\end{pmatrix},
\end{equation}
where the relative phase \( \theta \) is determined from
\begin{equation}
e^{ i \theta } =\frac{C_1}{|C_1|}=
\frac{
e^{ -i \xi \rho} \left[
J_{0}\!\left( \zeta \right)
e^{ -i (\rho-\eta) }
-
J_{2}\!\left( \zeta \right)
e^{ i (\rho - \eta) }
\right]
}{
\sqrt{
J_0^2\!\left( \zeta\right)
+
J_2^2\!\left( \zeta \right)
-
2 J_0(\zeta) J_2(\zeta)
\cos(2(\rho - \eta))
}
}.
\end{equation}

Here it is important to emphasize that both the quasienergies
and the quasi-eigenstates acquire a dependence on $\beta$, $\Delta$ and
 $\zeta$ through the Bessel functions. This dependence would be absent from $\varepsilon_{\mathrm{eff},\pm}$,
and most importantly from $\ket{u_{\pm}}$, were it not for the use of
the Magnus expansion in Eqs.~\eqref{eq:magnus01}–\eqref{eq:magnus03}
to approximate the unitary evolution operator.
In contrast, standard time-dependent perturbation theory generally
leads to a non-unitary approximation of the evolution operator,
which remains independent of $\beta$, $\Delta$ and
 $\zeta$.


\section{Micromotion and Envelope Dynamics}\label{sec:micromotion}

While the effective Hamiltonian is obtained from the stroboscopic
Floquet--Magnus expansion, the micromotion operator
can be derived by performing a Fourier expansion of the kick operator.
This procedure allows for an explicit separation between macromotion
and micromotion contributions in time-resolved observables.
To characterize the dynamics, we compute the time evolution of the band occupation probability using Eqs. (\ref{eq:Uz}) and (\ref{eq:Uphii}), yielding
\begin{equation}\label{eq:Pt}
  P_n(t) = \langle \chi(0)|\,\hat{U}_\chi^\dagger(t)\,\hat{P}_n\,\hat{U}_\chi(t)\,|\chi(0)\rangle,
\end{equation}
where $\hat{P}_n = |n\rangle\langle n|$ is the projector onto band $n=1,2$.
For pedagogical purposes, the following analysis is carried out using
projection operators; however, the formalism is fully general
and applies to arbitrary observables.
The time-evolution operator in the $\chi$ representation is given by
\begin{equation}\label{eq:Uchi}
  \hat{U}_\chi(t) = \hat{U}_z^\dagger(t)\,\hat{U}_\phi(\Omega t + \rho),
\end{equation}
where $\hat{U}_\phi$ is evaluated at the shifted variable $t' = \Omega t + \rho$,
with the initial condition $|\phi(\rho)\rangle = \hat{U}_z(0)|\chi(0)\rangle$
corresponding to physical time $t = 0$.
Expressing $\hat{U}_\phi(t')$ in terms of the effective Hamiltonian and
the micromotion operator,
\begin{equation}\label{eq:Uphi}
  \hat{U}_\phi(t') = e^{-i\hat{K}(t')}\,e^{-i\hat{H}_{\mathrm{eff}}\,t'}\,e^{i\hat{K}(0)},
\end{equation}
the dynamics decomposes into an initial kick $e^{i\hat{K}(0)}$,
a slow evolution governed by $\hat{H}_{\mathrm{eff}}$, and a periodic micromotion
described by $\hat{K}(t')$.  Fixing the gauge $\hat{K}(0)=0$ and substituting
into Eq.~\eqref{eq:Uchi}, the time-evolution operator becomes
\begin{equation}\label{eq:Uchi_full}
  \hat{U}_\chi(t)
  = \hat{U}_z^\dagger(t)\,
    e^{-i\hat{K}(\Omega t+\rho)}\,
    e^{-i\hat{H}_{\mathrm{eff}}\,\Omega t}\,
    e^{-i\hat{H}_{\mathrm{eff}}\rho},
\end{equation}
where
\begin{equation}\label{eq:Knew}
 e^{-i\hat{H}_{\mathrm{eff}}\,\rho}   = e^{i\hat{K}_{\mathrm{new}}(0)} 
\end{equation}
absorbs the phase accumulated into a
redefinition of the initial state.
The rotated initial state reads
\begin{equation}\label{eq:ketrot}
  |\chi_{\mathrm{rot}}(0)\rangle
  = e^{i\hat{K}_{\mathrm{new}}(0)}\,|\chi(0)\rangle.
\end{equation}
In this way, the phase $\rho$ enters the dynamics
as an effective initial rotation, while the subsequent evolution is
governed by the effective Hamiltonian $\hat{H}_{\mathrm{eff}}$,
followed by the time-periodic micromotion operator $\hat{K}(\Omega t + \rho)$
and the frame transformation
$\hat{U}_z^\dagger\!\left(t\right)$.

To separate the macromotion from the micromotion in Eq.~(\ref{eq:Pt}), we first apply the kick operator $\hat{K}(t')$ to the band occupation operator $\hat{P}_n$ using the Baker–Campbell–Hausdorff formula. This yields
\begin{equation}
e^{i\hat K(t^\prime)} \hat P_n e^{-i\hat K(t^\prime)}
=
\hat P_n+
 i [\hat K(t^\prime), \hat P_n]
- \frac{1}{2}[\hat K(t^\prime),[\hat K(t),\hat P_n]]
+ \cdots
\end{equation}
Using a Fourier expansion of the form $\hat K(t^\prime)= \sum_{m\neq 0} \hat K_m \, e^{i m  t^\prime}$  and collecting terms with the same phase
$e^{i m  t^\prime}$, one obtains
\begin{equation}
e^{i\hat K(t^\prime)} \hat P_n e^{-i\hat K(t^\prime)}
=
\hat P_n+ \sum_{m\neq0} \hat O^{(n)}_m e^{i m t^\prime}
\end{equation}
where
\begin{equation}
\hat O^{(n)}_m
=
i [\hat K_m, \hat P_n]
- \frac{1}{2}
\sum_{l}
[\hat K_m,[\hat K_{m-l},\hat P_n]]
+ \cdots \, .
\end{equation}
encodes the micromotion-induced harmonics at multiples of the driving frequency.
Therefore, using the equations $(\ref{eq:Pt})$, $(\ref{eq:Uchi})$ and $(\ref{eq:ketrot})$, $ P_n(t)$ takes the form 
\begin{multline} 
P_n(t)= \bra{\chi_{rot}(0)} e^{ i \hat{H}_{\mathrm{eff}} t}\hat U_z\!\left(t\right)\hat P_n 
U_z^{\dagger}\!\left( t \right) \,e^{- i \hat{H}_\mathrm{eff} t } \,
 \ket{\chi_{rot}(0)} \\
 +  \sum_{m \neq 0}
e^{i m  \left( \Omega t+\rho \right)}
\,
\bra{\chi_{rot}(0)}
e^{ i \hat{H}_{\mathrm{eff}} t}\hat U_z\!\left(t\right) \hat O^{(n)}_m 
U_z^{\dagger}\!\left(t\right) \,e^{- i \hat{H}_\mathrm{eff} t }
 \ket{\chi_{rot}(0)},\, \\
 = P_n^{(\mathrm{macro})}(t) + P_n^{(\mathrm{micro})}(t)
\end{multline}
where $P_n^{(\mathrm{macro})}(t)$ determines the slow envelope of
the dynamics  and $P_n^{(\mathrm{micro})}(t)$ describes the fast
oscillatory corrections around this
envelope, where the macromotion evolution is explicitly given by
\begin{equation}\label{eq:Pmacro}
     P_n^{(\mathrm{macro})}(t)=\bra{\chi_{rot}(0)} e^{ i \hat{H}_{\mathrm{eff}} t}\hat P_n  \,e^{- i \hat{H}_\mathrm{eff} t }
 \ket{\chi_{rot}(0)} \, .
\end{equation}
In the previous equation, the operator $\hat U_z(t)$ does not appear because the projector operator  of the valence and conduction bands commutes with $\hat \sigma_z$.

\section{Results and discusion}\label{sec:results}
In what follows, we restrict our analysis to the
macromotion of the occupation probability determined
by the of Eq. (\ref{eq:Pmacro}) with $\ket{\boldsymbol{\chi}(0)}=(1,0)^{T}$ given by
\begin{eqnarray}
P_1^{macro}(t) &=& \cos^2 (\omega_\mathrm{eff} t+\varphi),
\label{eq:P1macro}\\
P_2^{macro}(t) &=& \sin^2 (\omega_\mathrm{eff} t+\varphi),
\label{eq:P2macro}
\end{eqnarray}
where $\omega_\mathrm{eff}=\epsilon_{\mathrm{eff},+}-\epsilon_{\mathrm{eff},-}$
is the effective Rabi frequency and $\varphi= \omega_\mathrm{eff} \rho$.
We note that both $\omega_\mathrm{eff}$ and $\varphi$ do not depend on
the valley index; therefore, this result is valley-independent
and is governed by the quasienergy, Eq.(\ref{eq:energy}),
which depends on  the parameters of the electromagnetic wave: the frequency and the 
electric field intensity given by $\zeta$,
the polarization and helicity described by $\beta$,
and the relative angle $\Delta$ between the polarization direction
and the electron momentum.
In this section, all 
calculations are carried out using the following parameter values: 
$E_0 = 5 \times 10^{-2}\,\text{V/m}$, the electromagnetic wave frequency 
$\Omega = 22\,\text{GHz}$, and the corresponding strength parameter 
$\lambda = 0.156821$, corresponding to $\zeta \ll 1$, i.e., the 
weak-field (perturbative) regime.
In this limit, only the lowest Floquet harmonics contribute
significantly to the dynamics, justifying the  dominance of the $n=1$
resonant channel. Time is expressed in units of $\tau_1 = 2\pi/ \lambda$,
which sets the natural timescale of the envelope dynamics.

For clarity, we define the time evolution of the valence band
population as $P_1^{\mathrm{ana}}(t) = P_1^{\mathrm{macro}}(t)$.
In Figure (\ref{Fig1}), we compare the analytical result
$P_1^{ana}(t)$ with $P_1^{num}(t)$, obtained from  the numerical
solution of the coupled system of differential equations  Eq. 
(\ref{eq:DiracEquationInteraction}) as a funtion of $\tau_1$.
In panel (a), we present the time evolution of $P_1(t)$ for $\Delta=\pi/4$,
the black (red) solid line represents the numerical solution
for $\beta=1(-1)$ and the  black (red) dashed lines indicates
the analytical solution.
The dynamics exhibits rapid oscillations with pronounced
maxima approaching unity and narrow minima.
The analytical solution captures the slowly varying envelope
of the full dynamics, while the fast oscillatory structure arises
from the underlying micromotion.
In addition, we note that the polarization introduces a shift in the
time evolution of $P^{num/ana}_1(t)$ to the left (right) for $\beta = 1$
($\beta = -1$) at $\Delta = \pi/4$ (black and red lines, respectively).
This shift is due to the phase-induced initial Floquet kick
$\varphi$ given by Eq. (\ref{eq:ketrot}).
To quantitatively assess the validity of the macroscopic approximation 
$P_1^{ana}(t)$, we computed the root-mean-square error (RMSE) 
with respect to the full numerical solution $P_1^{num}(t)$ over 
the considered time interval. The RMSE is defined as
\begin{equation}
\mathrm{RMSE}
= \sqrt{
\frac{1}{N}
\sum_{n=1}^{N}
\left[
P^{ana}(t_n)
-
P^{num}(t_n)
\right]^2
} \times 100\%,
\end{equation}
where $t_n$ denotes the discrete time points used in the comparison 
and $N$ is the total number of sampled points and
$P^{ana/num}(t_n)=\int_{n \tau} ^{(n+1)\tau} P^{ana/num}(t) d t$
where $\tau=2\pi/\Omega$.
We found $RMSE=1.26\%$ for $N=100$ and $RMSE=5.11\%$ for $N=200$.
This behavior is understandable since the effective Hamiltonian
is obtained at zero order; however, if one wants to maintain
a more precise approximation over longer times, corrections of
higher order in the Magnus expansion must be included.
It is important to note that an error of $1.26\%$ over
100 periods represents a remarkably good approximation
for practical purposes.
In panel (b), we show the time evolution of the occupation probability 
$P_1(t)$ for $\beta=0,0.5,1$ fixing $\Delta=\pi/7$. 
The solid lines correspond to the full numerical solution 
$P_1^{num}(t)$, while dashed lines represent the
analytical solution $P_1^{ana}(t)$ as a function of  $\tau_1$.
We note that as $\beta$ decreases from $1$ to $0$,
the oscillation period of the envelope increases.
In the linear polarization limit $\beta = 0$ (blue curves),
the dynamics slows down significantly and the oscillations
become broader, reflecting the reduction of the effective Rabi
frequency in comparision with circular $\beta=1$ (black lines)
and elliptical $\beta=0.5$ (red curves) polarization.

In Figure \ref{Fig2} (a), we show the dependence of the effective Rabi frequency
$\omega_{\mathrm{eff}}(\Delta)$ for different values of $\beta$.
For circular polarization ($\beta = 1$, black dashed line), 
$\omega_{\mathrm{eff}}$ is constant therefore is independent of the
relative phase $\Delta$.
As $\beta$ decreases toward the linear polarization 
($\beta = 0$), a pronounced $\Delta$-dependence emerges, with  
minima and maxima values at $\Delta = 0,\pi,2\pi$ and 
$\Delta = \pi/2, 3\pi/2$, respectively.
The condition $\omega_{\mathrm{eff}}(\Delta)=0$
indicates the suppression of the resonant interband coupling,
so that no net population transfer occurs at the stroboscopic level,
for this case the dynamics is then entirely governed by micromotion,
consistent with earlier reports in driven Dirac systems \cite{alphaT3}.
As $\beta$ decreases, the variation of $\omega_{\mathrm{eff}}$ with
$\Delta$ becomes more pronounced, leading to deeper minima and larger
contrast between extrema as a result of stronger anisotropic interference effects.
In Figure \ref{Fig2} (b), we present a contour plot of
$\omega_{\mathrm{eff}}(\Delta,\beta)$.
The structure reveals symmetric lobes with $\pi$-periodicity
in $\Delta$, consistent with the angular dependence of the quasienergy 
splitting.

The behavior of $\omega_{\mathrm{eff}}(\Delta,\beta)$
can be understood from the effective quasienergy
in Eq.~(\ref{eq:energy}), where the angular dependence
is encoded in the interference term proportional
to $\cos(2\rho - \eta)$.
The factor $2\rho$ explains the observed $\pi$-periodicity in $\Delta$. 
The extrema values of $\omega_{\mathrm{eff}}$ occur when 
$\cos(2\rho-\eta)=\pm 1$. 
In particular, maxima arise for $\cos(2\rho-\eta)=-1$, 
corresponding to constructive interference 
$\propto |J_0+J_2|$, whereas minima occur for 
$\cos(2\rho-\eta)=+1$, yielding destructive interference 
$\propto |J_0-J_2|$. For circular polarization ($\beta = 1$),
we find that $C_1$ (see Eq.~(\ref{eq:C1})) is proportional to
$e^{i\Delta}$, so that $\epsilon_{\text{eff},\pm}$ is independent
of $\Delta$, in agreement with the flat behavior observed
in Fig.~\ref{Fig2}(a).
Conversely, near the 
linear polarization regime ($\beta \approx 0$), the interference term 
dominates, leading to strong angular modulation and the pronounced 
lobular structure visible in Figure \ref{Fig2} (b).
In Figure (\ref{Fig3}), we show the phase $\varphi$ as a function of 
the relative angle $\Delta$ and the polarization parameter $\beta$.
Panel (a) shows $\varphi(\Delta)$ for representative values 
$\beta = 0.01, 0.2, 0.6, 1$. For $\beta < 1$, the phase exhibits 
sharp variations near $\Delta = \pi/2$ and $3\pi/2$, accompanied by 
sign reversals, whose origin will be discussed below. As $\beta$ 
increases, the angular dependence becomes progressively smoother and 
more symmetric. In the circular limit ($\beta = 1$, dashed line), the phase varies 
linearly with $\Delta$ within each half-period, with abrupt sign 
reversals at $\Delta = \pi/2$ and $3\pi/2$. As $\beta$ decreases, 
the linear behavior is lost and the phase develops an increasingly 
asymmetric profile and change of sign within each half-period.

Panel (b) 
displays the full $(\Delta, \beta)$ map, where the phase exhibits a 
clear $\pi$-periodicity in $\Delta$ with alternating positive and 
negative regions separated by nodal lines. The angular structure is 
most pronounced for small $\beta$ and gradually weakens as 
$\beta \rightarrow 1$, consistent with the restoration of rotational 
symmetry under circular polarization. The behavior of 
$\varphi(\beta, \Delta)$ can be understood from the explicit 
expressions for $\omega_{\mathrm{eff}}$ and $\rho$ given by 
Eqs.~(\ref{eq:energy}) and (\ref{eq:functions}). The sharp features 
near $\Delta = \pi/2$ and $\Delta = 3\pi/2$ arise from the rapid 
variation of $\rho$ as $\tan\Delta$ diverges. Since 
$\omega_{\mathrm{eff}} > 0$ for all values of $\Delta$, as shown in 
Figure ~(\ref{Fig2}), the sign change of $\varphi$ is entirely 
determined by the sign of $\rho$. This manifests in $P_1(t)$ as a 
phase shift of the oscillations: for $\beta > 0$, the shift occurs 
to the left (right) for $\Delta \in [0, \pi/2]$ 
($\Delta \in [\pi/2, \pi]$) whereas for $\beta < 0$ the opposite behavior occurs.

\section{Conclusions}\label{sec:conclusions}
We have developed a resonant Floquet-Magnus framework to describe
Dirac electrons in graphene driven by polarized electromagnetic radiation.
The analysis was performed in the interaction picture, where a unitary
transformation analogous to a rotating-wave approximation was used to
remove the diagonal terms of the interaction Hamiltonian.
This procedure yields a generalized Rabi-type Hamiltonian whose matrix
elements are complex time dependent. Within this formulation,
the Floquet-Magnus expansion allows us to identify the resonance condition.
By imposing this condition, the stroboscopic time-averaging procedure isolates
a single dominant Floquet harmonic, leading to an effective two-level
Hamiltonian that governs the macromotion of the system.
In addition, we find that the evolution operator acquires an initial
unitary factor of the form $e^{-i \hat{H}_{\mathrm{eff}} \rho}$,
which depends on the effective Hamiltonian, the polarization
parameter $\beta$, and the relative angle $\Delta$ between the
electron momentum and the polarization direction.
This factor acts as a resonant initial Floquet kick, redefining the
initial state of the system. Through a Fourier analysis of the
time-evolution operator further enables a clear separation between
macromotion and micromotion contributions.
Although the derivation of $\hat{H}_{\mathrm{eff}}$,
the initial kick, and the macro/micromotion decomposition has
been carried out under the resonance condition, the methodology can
be systematically extended to other Dirac materials.
In the resonant regime, the quasienergy splitting depends
nontrivially on both the $\beta$ and $\Delta$.
Circular polarization restores rotational symmetry in momentum
space and yields an angle-independent effective Rabi frequency.
In contrast, elliptical and linear polarizations introduce
anisotropic interference effects arising from competing Bessel harmonics,
producing a characteristic $\pi$-periodic angular modulation.
Comparison with full numerical simulations shows that the zeroth-order
Floquet-Magnus approximation accurately reproduces the envelope dynamics
in the resonant regime, while deviations originate from micromotion
and higher-order corrections.
These results demonstrate that the polarization parameters
$\beta$ and $\Delta$ constitute independent and tunable control knobs
for engineering effective couplings and initial states
in periodically driven Dirac systems in the resonant regime.
The $\beta$- and $\Delta$-dependence of the quasi-eigenstates becomes apparent
only through the use of the Magnus expansion of the evolution operator;
otherwise, one would obtain $\beta$- and $\Delta$-independent expressions.
These results suggest a clear experimental route to demonstrate
the tunability of the electronic states in driven graphene through
angle-resolved photocurrent measurements.
In particular, the directional dependence of the photocurrent
should reflect the control exerted by the polarization
parameters $\beta$ and $\Delta$, and a Kubo-formula analysis
would provide a natural framework to quantify the anisotropic
conductivity underlying the magnitude and direction of the
photoinduced current.
A detailed Kubo-formula analysis of the anisotropic photocurrent
response will be addressed in future work.
\section{Acknowledgements}\label{sec:Acknowledgements}
This work was supported by DCB UAM-A grant numbers
22322035, and 22322036.\\


\end{document}